\documentclass[12pt]{article}
\usepackage{amsmath}
\usepackage{graphicx}
\usepackage{natbib}
\usepackage{url} 

\usepackage{hyperref}
\usepackage{algorithm}
\usepackage{algorithmicx}
\usepackage{algpseudocodex}
\usepackage{amsfonts}
\usepackage{float}

\usepackage{amsthm}

\graphicspath{{figures/}}

\newcommand{\blind}{0}

\begin{document}

\def\spacingset#1{\renewcommand{\baselinestretch}%
{#1}\small\normalsize} \spacingset{1}


\if0\blind
{
  \title{\bf DRtool: An Interactive Tool for Analyzing High-Dimensional Clusterings}
  \author{Justin Lin \\
    Department of Mathematics, Indiana University \\
    ORCiD: 0009-0007-7190-2430 \\
    linjus@iu.edu \\
    and \\
    Julia Fukuyama \\
    Department of Statistics, Indiana University \\
    ORCiD: 0000-0002-7590-5563}
   \date{}
  \maketitle
} \fi

\if1\blind
{
  \bigskip
  \bigskip
  \bigskip
  \begin{center}
    {\LARGE\bf Title}
\end{center}
  \medskip
} \fi

\bigskip
\begin{abstract}
When faced with new data, we often conduct a cluster analysis to obtain a better understanding of the data's structure and the archetypical samples present in the data. However, the increases in data complexity and dimensionality have made this step very tricky. The large proportion of noise in high-dimensional data blurs patterns and trends, making clusters difficult to distinguish. As such, cluster-discovery tools and cluster-verification tools must be adapted to address the difficulties of high-dimensional data. Nonlinear dimension reduction is a step in the right direction, but even these methods are known to produce false structures, especially when mishandled. A common phenomenon that often goes undetected by the untrained eye is over-clustering of the data. In continuation of these efforts, we developed new cluster verification techniques, including visual assessments and a hypothesis test, that help analysts distinguish false clusters and better interpret their high-dimensional clustering results. For ease of use, these new methods are provided in an interactive toolbox available via R package \href{https://www.github.com/justinmlin/DRtool}{DRtool}.
\end{abstract}

\noindent%
\vfill

\newpage
\spacingset{1.75} 

\section{Introduction}
High-dimensional cluster analysis is notoriously difficult due to the inconsistency and noisiness of high-dimensional distances \citep{high_dim_cluster}. The unreliability of these distances makes the application of classical clustering methods to high-dimensional data specious at best. 

Data visualization goes hand-in-hand with cluster analysis because it provides means for both cluster discovery and cluster verification. Nonlinear dimension reduction methods are being developed to handle and visualize high-dimensional data, but these methods are not without faults. The potency of nonlinear dimension reduction methods lies in their flexibility, allowing them to model complex data structures. That same flexibility, however, makes them difficult to use and interpret. Each method requires a slew of hyperparameters that need to be calibrated, and even when adequately calibrated, these methods require a trained eye to interpret. For example, the two most popular nonlinear dimension reduction methods, t-SNE and UMAP, sometimes generate misleading results \citep{understanding_UMAP, Distill}. The results often cluster, even when no clusters exist in the data, and cluster sizes/locations can be unreliable. This is especially problematic in high-dimensional cluster analysis when the tools for determining cluster legitimacy are limited. 

With this in mind, we have developed new cluster-verification methods tailored to high-dimensional data. These methods are conveniently bundled in an interactive toolbox analysts may use to conduct a post-hoc analysis of their high-dimensional clustering. These methods use the minimum spanning tree (MST) to provide a broad view of the data's global structure as well as local perspectives on inter-cluster relationships. When dealing with highly complex data, the convenience and variety of a complete toolbox allows analysts to extract more information from their high-dimensional cluster analyses.

In this paper, we discuss related works addressing a similar problem (Section 2). We describe the analytical plots provided by the tool (Section 3). We present an MST stability experiment, demonstrating the MST's ability to approximate high-dimensional structure, as well as power and size analyses for a novel hypothesis test (Section 4). And we walk through the use of the tool on two separate data sets (Section 5).

\section{Related Works}
Much work has gone into assessing cluster analyses, both quantitatively and visually. Quantitative metrics can only convey information numerically, so their extent is useful but rather limited. For example, the gap statistic \citep{gap} and average silhouette score \citep{silhouette} are both numerical measures of cluster performance mainly used to determine the optimal number of clusters, but they do not convey any information about the individual clusters. Comparatively, visual assessments have larger scopes and come in many different forms.

Classical cluster visualization methods (see \citet{cluster_visualization_review} for a review) provide a view of the global cluster relations. The dendrogram in hierarchical clustering provides a visual description of how the samples are sequentially merged (agglomerative clustering) or split (divisive clustering). Its tree-like structure encapsulates the general proximity of clusters. A plot of the per-sample silhouette scores \citep{silhouette} provides a visual indication of cluster separation and overlap. Convex cluster hulls, like those implemented in \cite{ClusterHulls}, can be used to visualize approximated cluster perimeters but ignore cluster densities. These methods provide a general sense of global inter-cluster proximities but 
fail to capture local inter-cluster relationships.

The Grand Tour \citep{grand_tour} and its modern variations \citep{tour_review} provide a more interactive view of high-dimensional data. These methods aim to visualize high-dimensional data through a sequence of linear projections and often include interactive elements like pausing or reversing the sequence of images. While the interactive and multi-perspective view is very powerful, the linear projections are susceptible to over-crowding, often obscuring local structure \citep{burning_sage}. This is why nonlinear dimension reduction methods are preferred when it comes to high-dimensional clustering.

\citet{choosing_NLDR} take the tour method and applies it to nonlinear dimension reduction. The method fits a triangulation to a nonlinear embedding of the data, lifts the triangulation to high dimension, then overlaps the triangulation on each projection in the tour. The tour provides a multi-perspective view of the triangulation's fit, indicating how faithful the nonlinear embedding is to the high-dimensional data. While this method provides a visual assessment of nonlinear dimension reduction, like the other previously discussed methods, it only provides a macro-level view of the clusters.

One method, capable of visualizing micro-level or pairwise inter-cluster relationships, is shadow-stars \citep{shadow_stars}. Shadow values are an adaptation of the usual silhouette score, but instead of comparing within-cluster distances to out-of-cluster distances, shadow values compare distance to the nearest cluster to distance to the second nearest cluster. The pairwise nature of these shadow values allows the user to study the relationships between adjacent clusters via various plots. Like silhouette scores, however, shadow values only represent one perspective.

DRtool builds upon these ideas by providing an interactive and highly customizable toolbox. It presents the user with various analytical plots and tests, providing both a global view of the inter-cluster structure and local pairwise relationships. It also includes an interactive interface and an increased level of customization, allowing the user to readily select custom clusters or overlay additional information like kernel density estimates.

The use of graphs, and specifically the MST, to model high-dimensional data has also been well-documented and is explained in Section 3.1.

\section{Methods}

Given a clustering of high-dimensional data, DRtool uses the MST to extract and analyze the inter-cluster relationships.

\subsection{Minimum Spanning Tree}
Graphs have been applied to many multivariate statistical problems. The authors of \citet{MAP_test} introduced the minimal ascending path spanning tree as a way to test for multimodality. The Friedman-Rafsky test \citep{Friedman-Rafsky-test}, along with its modern variations \citep{Friedman-Rafsky1, Friedman-Rafsky2, Friedman-Rafsky3}, uses the MST to construct a multivariate two-sample test. Single-linkage clustering \citep{single-linkage} and runt pruning \citep{runt_pruning} are both intimately related to the MST. In the context of dimension reduction, IsoMap \citep{IsoMap} makes use of neighborhood graphs, \citet{MIST_example} introduce the maximum information spanning tree, and \citet{MST_example} use the MST. These methods, which fall under the category of manifold learning, use graphs to model high-dimensional data assumed to be drawn uniformly from a high-dimensional manifold. An accurate low-dimensional embedding can then be constructed from these graphs. It's apparent that graphs are useful for describing high-dimensional data, especially when it comes to dimension reduction and cluster analysis. Our tool uses the MST to analyze the reliability of visualizations produced by nonlinear dimension reduction methods.

We've opted for the MST for a couple of key properties. Firstly, the MST and shortest paths along it are quick to compute. Secondly, the MST contains a unique path between any two vertices, providing a well-defined metric on the data. Lastly, it provides a good summary of the data's structure. It contains as a subgraph the nearest-neighbor graph, and any edge deletion in the MST partitions the vertices into two sets for which the deleted edge is the shortest distance between them \citep{Friedman-Rafsky-test}.

\subsubsection{Simplified Medoid Subtree}
The MST is meant to provide a robust description of the data's inter-cluster relationships. So while the MST may be locally affected by noise, we only require the representation of inter-cluster relationships to be stable in the presence of noise. To demonstrate MST stability, we study the effect of random noise on the inter-cluster relationships explained by the MST.

\begin{figure}[H]
\begin{center}
\includegraphics[width=6in]{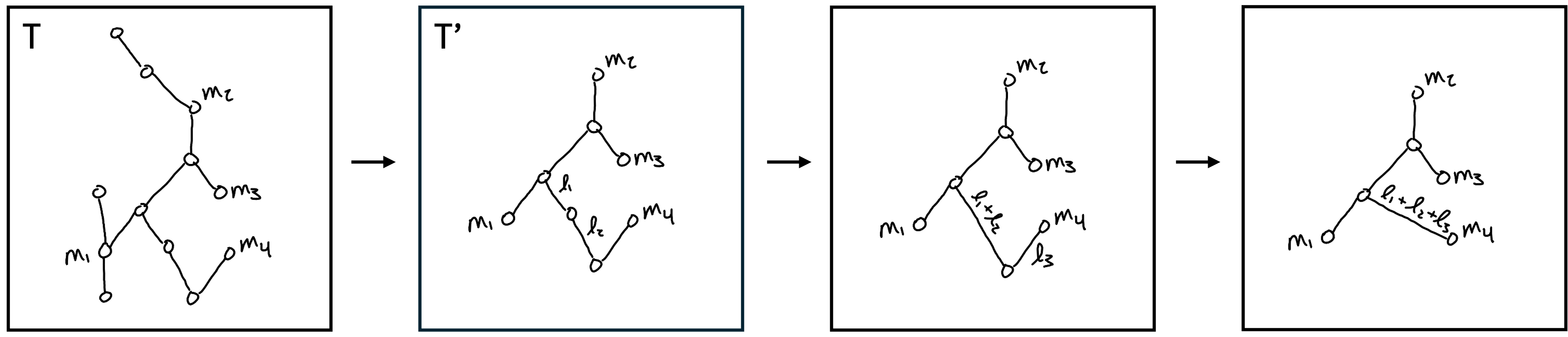}
\end{center}
\caption{Example of Algorithm \ref{alg1}. The first step is to compute the medoid subtree $T'$. Then non-medoid degree-2 vertices are replaced with single edges.}
\label{alg1 ex}
\end{figure}

To encode the inter-cluster relationships present in the MST, we define the simplified medoid subtree as follows. We take the medoid subtree, i.e. the minimal subtree containing the medoid of each cluster, then apply a simplification procedure (Algorithm \ref{alg1}). The algorithm collapses paths of non-medoid vertices into single edges of equal length by iteratively replacing degree-2 vertices with a single edge. We refer to the output as the simplified medoid subtree. It encodes the global inter-cluster relationships within the data. See Figure \ref{alg1 ex} for an example.

\begin{algorithm}[H]
\caption{Simplified Medoid Subtree}
\begin{algorithmic}[1]
\Require MST $T = (V, E)$ with vertex set $V$, edge set $E$,  and cluster medoids $m_1, \hdots, m_k \in V$
\State $T' = (V', E') \Leftarrow$ minimal subtree of $T$ containing all $m_i$
\Repeat
	\State Let $v \in V' \setminus \{m_1,  \hdots, m_k\}$ with degree $deg(v) = 2$ and neighbors $a, b \in V'$. Let $d(v, a)$ and $d(v, b)$ be the weights of the edges incident to $v$.
	\State Replace $v$ and its two incident edges with an edge connecting $a$ and $b$ with weight $d(v, a) + d(v, b)$.
\Until{$T'$ no longer contains non-medoid vertices with degree two.}
\State \Output T'
\end{algorithmic}
\label{alg1}
\end{algorithm}

\subsubsection{Robinson-Foulds Metric}
To compare simplified medoid subtrees, we use the Robinson-Foulds metric \citep{RF_metric}. The R-F metric was originally introduced to quantify the dissimilarity of phylogenetic trees, but the algorithm generalizes to arbitrary weighted trees. It looks at partitions of each tree created by removing individual edges then counts the number of partitions present in one tree but not the other. We modified the R-F metric to specifically measure the dissimilarity in partitions of the medoid vertices. In particular, Algorithm \ref{alg2} compares the partitions of the medoid vertices created by removing each individual edge.

\begin{algorithm}[H]
\caption{Robinson-Foulds Distance}
\begin{algorithmic}[2]
\Require Trees $T_1 = (V_1,E_1)$ and $T_2 = (V_2, E_2)$ with medoids $m_1, \hdots, m_k \in V_1$ and $n_1, \hdots, n_k \in V_2$
\State $P_1 \Leftarrow \{\}$
\For{$e \in E_1$}
	\State $G \Leftarrow (V_1, E_1 \setminus \{e\})$ with components $G_1$ and $G_2$
	\State $M_1 \Leftarrow \{m_1,\hdots,m_k\} \cap V(G_1)$
	\State $M_2 \Leftarrow \{m_1,\hdots,m_k\} \cap V(G_2)$
	\State $P_1 \Leftarrow \Call{Add}{P_1, \{M_1, M_2\}}$
\EndFor
\State $P_2 \Leftarrow \{\}$
\For{$e \in E_2$}
	\State $G \Leftarrow (V_2, E_2 \setminus \{e\})$ with components $G_1$ and $G_2$
	\State $M_1 \Leftarrow \{n_1,\hdots,n_k\} \cap V(G_1)$
	\State $M_2 \Leftarrow \{n_1,\hdots,n_k\} \cap V(G_2)$
	\State $P_2 \Leftarrow \Call{Add}{P_2, \{M_1, M_2\}}$
\EndFor
\State \Output $\frac{\left|P_1 \Delta P_2 \right|}{2\left| P_1 \cap P_2 \right|}$
\end{algorithmic}
\label{alg2}
\end{algorithm}

\subsection{The Tool}
The main objective is to analyze and leverage the structural data embedded in the MST. For example, paths between clusters are used to study inter-cluster relationships in the context of the underlying manifold from which the data are assumed to be drawn.

To start, the user must provide a data matrix, a low-dimensional embedding, and a clustering. From there, the MST is calculated and various analytical plots are provided. The primary plot is the low-dimensional embedding colored according to the provided clustering. There is an option to overlay the medoid MST to understand the global structure of the clusters.

The remaining plots require the user to select two groups of interest, which can be done interactively in one of two ways. One way is to select two endpoints. The MST path is calculated and projected onto the low-dimensional embedding. The two groups are then the classes each endpoint belongs to. The second way is to select custom groups. The user may interact with the low-dimensional embedding by drawing boundaries for each group. The projected path then connects the medoid of each group. Once the groups and path are specified, the user is provided additional plots used to investigate the relationship between the two selected groups of points.

From now on, ``group" will refer to the user-selected groups, while ``cluster" will refer to the clusters in the user-inputted clustering.

\subsection{Path Projection Plot}
To better understand the path of interest, a local projection method is applied to visualize the path and nearby points in two dimensions. The goal of the projection is to ``unwind" the path, so it can be used to study the relationship between the two selected groups. We apply Principal Component Analysis followed by regularized Canonical Correlation Analysis in a method we've dubbed the PCA -- RCCA method.

\subsubsection{The PCA -- RCCA Method}
Let $p_1, \hdots, p_k \in \mathbb{R}^p$ be the points along the path and
$$P = \begin{bmatrix}
\text{---} & p_1 & \text{---} \\
 & \vdots & \\
\text{---} & p_k & \text{---}
\end{bmatrix} \in \mathbb{R}^{k \times p}$$
the matrix of path points.

The concept is to use Canonical Correlation Analysis to determine the two-dimensional linear projection that best unwinds the path. Given two matrices, CCA iteratively calculates linear combinations of the matrix variates for each matrix, known as canonical variate pairs, that maximize covariance. These pairs are chosen to be orthogonal, so they give rise to a projection subspace. To unwind $P$, we use CCA to compare $P$ against a degree $s$ polynomial design matrix $P_s$, $$P_s = \begin{bmatrix}
1 & 1^2 & \cdots & 1^s \\
2 & 2^2 & \cdots & 2^s \\
\vdots & \vdots & & \vdots \\
k & k^2 & \cdots & k^s
\end{bmatrix}.$$
In doing so, we are linearly transforming the path so that its shape best matches that of a degree $s$ parametric polynomial. The first two canonical variate pairs are used to construct a two-dimensional projection that maximizes the covariance between the projections of $P$ and $P_s$. This process generates a two-dimensional projection subspace onto which we can project the path and both groups. 

In most cases, however, regularization is required to avoid singularity because $p$ is often much greater than $k$. RCCA \citep{RCCA} adds a regularization constant along the diagonals of each covariance matrix to ensure nonsingularity. The regularization constant for $P$ is chosen via cross-validation, and no regularization constant is needed for $P_s$.

One issue with this method is the projected path often travels along the outskirts of the plot. This is due to the near-orthogonality of high-dimensional data \citep{near-orthogonal}. Because the non-path points are often nearly orthogonal to the projection subspace, they are overly shrunk in the projection. The path points are less affected because the projection subspace is selected to retain the path's shape. While this phenomenon doesn't discredit the entire plot, it leads to misrepresentation of the path's location relative to the rest of the points.

To alleviate this issue, we apply PCA on the entirety of $X$ prior to applying RCCA. Removing extraneous dimensions containing mostly noise limits the confusion of excess noise for independence. When RCCA is applied post-PCA, the projected path's relative position to the rest of the points is more credible.

\subsubsection{Calibrating Hyperparameters}
The user is responsible for calibrating the dimensionality of the PCA step and the degree $s$ of the reference polynomial design matrix. To pick a number of dimensions, the user is recommended to start with a moderately large number, relative to the dimensionality of the original data. The proportion of variance retained in the selected number of dimensions is conveniently displayed in the upper righthand corner of the plot. A larger number of dimensions retains more information but may misrepresent the location of the path relative to the rest of the points, while a smaller number of dimensions may diminish some of the variation in the data. As such, the user is encouraged to try different numbers of dimensions. To calibrate $s$, it is recommended to start with $s = 2$ then increment $s$ until the shape of the path stabilizes.

The user is also given the option to overlay a kernel density estimate. In order to do so, the bandwidth must be calibrated. The recommended procedure is to begin with a large bandwidth that estimates one mode, then gradually decrease the bandwidth until two modes appear. If the two modes correspond with the two groups of interest, and more modes do not immediately appear when continuing to decrease the bandwidth, then a bimodal distribution is a reasonable way to describe the data.

\subsection{The MST Test}
Another perspective on the relationship between the two selected groups can be gained from studying the local structure of the MST. The degree of connectivity between the two groups within the MST serves as a measure of separation. A large degree of connectivity indicates lesser separation, while a small degree of connectivity indicates more separation. This idea motivates a hypothesis test.

\subsubsection{The Test Statistic}
The test statistic, meant to quantify local connectivity, is based on the number of edges connecting the two groups of interest. However, counting single edges is too restrictive a measure. Consider the case when the two selected groups are polar ends of the same cluster. Because the medial region of the cluster does not belong to either group, there will be zero edges connecting the two groups, indicating the maximal degree of separation. This result is undesirable because the two groups actually belong to the same cluster in this hypothetical scenario.

\begin{figure}[H]
\begin{center}
\includegraphics[width=6in]{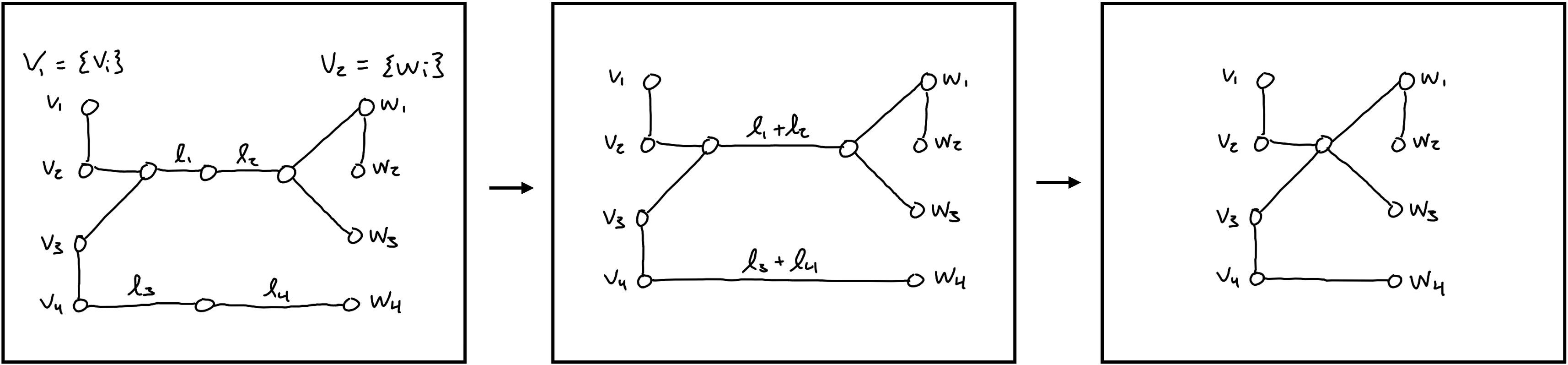}
\end{center}
\caption{Example of Algorithm \ref{alg3}. First, degree-2 vertices belonging to neither group are replaced with single edges. Then adjacent vertices belonging to neither group are collapsed into a single vertex.}
\label{alg3 ex}
\end{figure}

Instead, the test statistic counts the number of connecting paths rather than single edges. These paths are referred to as crossings and are counted according to the following procedure. The minimal subtree containing both groups is isolated. Because the two groups may not be adjacent in the MST, this subtree may contain points belonging to other clusters as well. To extract the structural relationship between the two groups of interest, the subtree must be simplified. The simplification process collapses paths between the two groups of interest into edges that can be counted (Algorithm \ref{alg3}). See Figure \ref{alg3 ex} for an example.

\begin{algorithm}[H]
\caption{Simplify Subtree}
\begin{algorithmic}[3]
\Require Tree $T = (V,E)$, group one vertices $V_1 \subset V$, and group two vertices $V_2 \subset V$
\State $T' = (V', E') \Leftarrow$ minimal subtree of $T$ containing $V_1 \cup V_2$
\Repeat
	\State Let $v \in V' \setminus (V_1 \cup V_2)$ with $deg(v) = 2$ and neighbors $a, b \in V'$. Let $d(v, a)$ and $d(v, b)$ be the weights of the edges incident to $v$.
	\State Replace $v$ and its two incident edges with an edge connecting $a$ and $b$ with weight $d(v, a) + d(v, b)$.
\Until{$T'$ no longer contains non-group vertices with degree two.}
\Repeat
	\State Let $v_1, v_2 \in V' \setminus (V_1 \cup V_2)$ be adjacent.
	\State Collapse the edge connecting $v_1$ and $v_2$. The combined vertex is adjacent to all neighbors of $v_1$ and $v_2$.
\Until{$T'$ no longer contains adjacent non-group vertices.}
\end{algorithmic}
\label{alg3}
\end{algorithm}

To count the number of crossings, the number of edges between the two groups in the simplified subtree is counted. It is also possible for a point of non-interest to act as a mediator along a path between the two groups of interest (see the unlabeled vertex in the third panel of Figure \ref{alg3 ex}). To account for this scenario, for each point of non-interest adjacent to both groups, we also count its maximal degree to both groups.

\subsubsection{The Null Distribution}
The null distribution should correspond to the number of crossings in the case when both groups belong to the same cluster. Because a cluster can be drawn from a number of unimodal distributions, we must consider a composite null hypothesis including all such distributions. Among these distributions, the distribution that maximizes the probability of rejection must be used to ensure the test has the correct size. That way, the probability of Type I error does not exceed the pre-specified significance level under any member of the composite null hypothesis.

We are in search of the unimodal distribution that minimizes the number of edges crossing a pre-specified hyperplane, representing the boundary between the groups. Finding this distribution is an intractable problem, so assumptions must be made. If we assume the number of edges crossing the hyperplane is proportional to the marginal density in a neighborhood around the hyperplane, then we may reduce the problem to the one-dimensional case.

Let $n_1, n_2 > 0$ be the sample sizes of each group and $c \in [-1, 1]$ the location of the mode. Let $\mathcal{F}$ be the family of distributions on $[-1, 1]$ such that
\begin{itemize}
	\item $f$ is increasing on $[-1, c]$,
	\item $f$ is decreasing on $[c, 1]$,
	\item $\int_{-1}^0 f = \frac{n_1}{n_1+n_2}$, and
	\item $\int_0^1 f = \frac{n_2}{n_1+n_2}$.
\end{itemize}
Let $\epsilon \in (0, 1)$. The aim is to find a $f \in \mathcal{F}$ that minimizes $\int_{-\epsilon}^\epsilon f$.

The proof (Appendix A) is broken up into multiple cases based on the values of the ratio $n_1/n_2$ and $c$. In all cases, a piecewise constant solution exists in which the density near the hyperplane is proportional to the density of the less dense group. This formal problem motivates the procedure by which the null distribution is simulated. Let $G_1 \in \mathbb{R}^{n_1 \times p}$ and $G_2 \in \mathbb{R}^{n_2 \times p}$ be the data matrices for each group. First, the density of each group is approximated, $$D_j = \frac{n_j}{\prod_i \sigma_i^j} \textrm{ for groups }j = 1,2$$ where $\sigma_1^j, \hdots, \sigma_{\min\{n_j,p\}}^j$ are the singular values of $G_j$.  The product of singular values is used to estimate the volume of each group because high-dimensional clusters tend to look Gaussian \citep{near-orthogonal}. Extraneous noise dimensions are removed prior to this process to avoid biasing the volume estimates. Now suppose $D_1 < D_2$. Then $n_1$ points are uniformly sampled from the hyperrectangle $$R = \prod_{i=1}^p \left[-\frac{\sqrt{12}}{2}\sigma_i^1, \frac{\sqrt{12}}{2}\sigma_i^1\right].$$ Note, $R$ has side lengths $\sqrt{12}\sigma_i^1$ so that the marginal distribution along the $i^\textrm{th}$ dimension has standard deviation $\sigma_i^1$, i.e. the standard deviation of the $i^\textrm{th}$ principal component of the original data. From there, the number of edges in the MST crossing the $x_1 = 0$ hyperplane is recorded. Repeated simulation of this procedure yields an approximate null distribution to which the test statistic is compared. The returned $p$-value is the percentile of the test statistic within this bootstrapped null distribution. A one-sided test is employed because we are only interested in rejecting the null for sufficiently small numbers of edge crossings.

See Section 4.2 for power and size analyses.

\subsection{Heatmap}
An interactive heatmap of the high-dimensional data is also supplied by the tool. A heatmap of the selected groups is a very powerful tool for comparison because it provides a feature-by-feature perspective of the high-dimensional data. The rows of the heatmap correspond to individual samples and are partitioned into the two selected groups. The columns correspond to features and are ordered by difference in group means. The heatmap highlights the exact features in which the two groups differ the most. It is also interactive, allowing the user to select and analyze sub-heatmaps, providing a more focused view on features of interest.

\subsection{Metadata Plot}
Along with the data and clustering, the user may also supply metadata corresponding to the samples in the original data. The metadata for each selected group is presented via pie charts for categorical data and box plots for numerical data, presenting a visual description of how the metadata differs between the two groups. These plots are especially useful for displaying categorical data not included in the numerical data set.

\section{Simulation Studies}

\subsection{MST Stability Experiment}
To demonstrate the MST's robust ability to describe global structure, we conducted a stability experiment. 1,500 samples were randomly chosen from the MNIST data set of handwritten digits \citep{MNIST}. Each $28 \times 28$-pixel image was flattened into a vector of length $28^2 = 784$, so the data contain 1,500 samples in $784$ dimensions. A PCA pre-processing step was employed to reduce the number of dimensions to 300. The simplified medoid subtree $T$ was then calculated.

Random Gaussian noise was then added to the data and the new simplified medoid subtree $T'$ was calculated. The R-F distance $RF(T, T')$ was recorded. This process was repeated 30 times.

\begin{figure}[H]
\begin{center}
\includegraphics[width=6in]{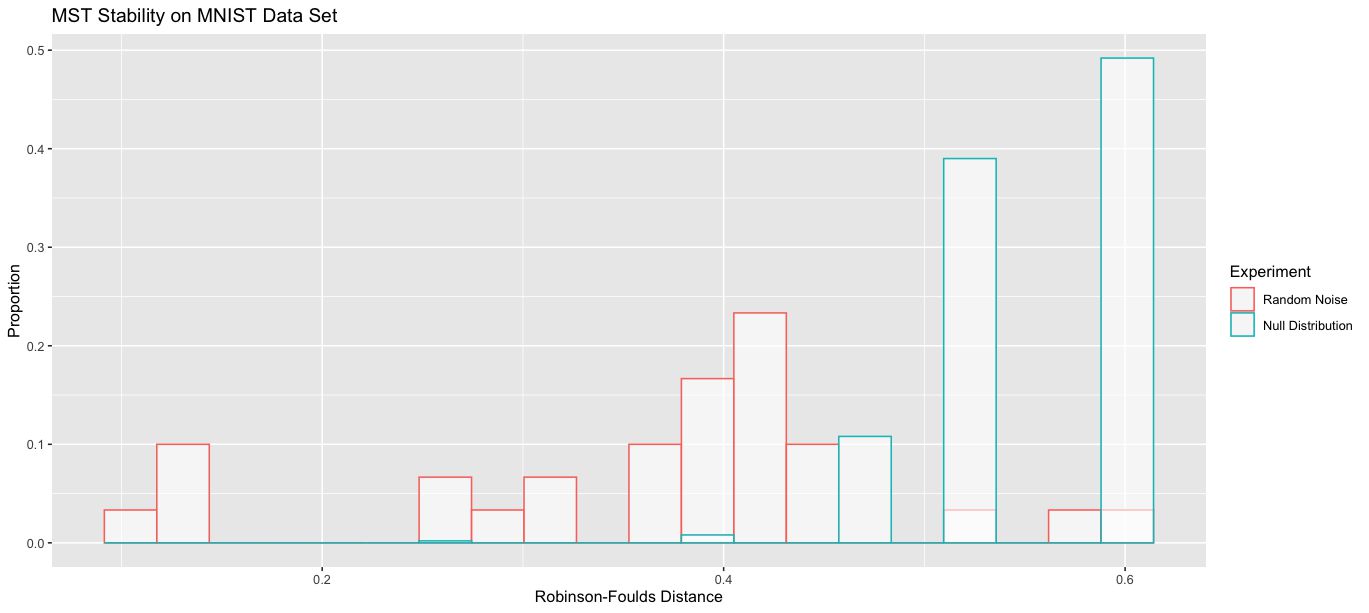}
\end{center}
\caption{MST stability results on the MNIST data set.}
\label{RF stability}
\end{figure}

To better interpret the R-F distances, we designed a null distribution of distances as a reference for comparison. These distances should represent R-F distances between trees that do not portray similar global structures and inter-cluster relationships. To generate the null distribution from the data, we randomly permuted the class labels and computed the R-F distances between the resulting simplified medoid subtrees and the original simplified medoid subtree. By randomly re-labelling the clusters, we are simulating examples with distinct global structures. Figure \ref{RF stability} shows the R-F distances produced by adding noise and permuting the class labels. The simplified medoid subtrees generated by adding noise were significantly closer to the original simplified medoid subtree than those generated by randomly permuting the class labels in terms of R-F distance, showing inter-cluster relationships in the MST are robust to noise.

\subsection{Power and Size Analyses of the MST Test}
The power of the test is dependent on the relative degree of separation between the two clusters. The experiment is set up as follows. Let $d \in (0,1]$. 50 points are randomly sampled from $[-2, -d] \times [-1,1]^{p-1}$, and 50 points are randomly sampled from $[d, 2] \times [-1,1]^{p-1}$. In other words, two hyperrectangular p-dimensional clusters are sampled and separated by a distance of $2d$. The MST test is run and the $p$-value is recorded. Through simulation, the power at varying levels of $d$ and $p$ is estimated. The power is expected to increase with $d$ and decrease with $p$. In high dimension, distances are inflated due to the increased noise-to-signal ratio, so the inter-cluster separation appears less significant.

\begin{figure}[H]
\begin{center}
\includegraphics[width=6in]{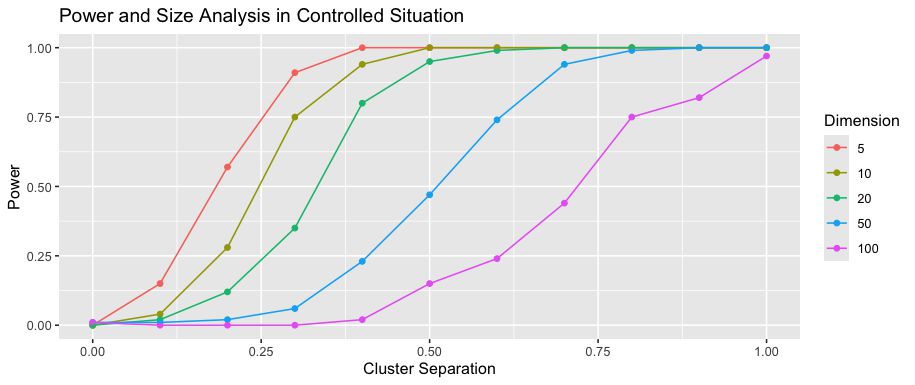}
\end{center}
\caption{Power analysis of the MST test.}
\label{power analysis}
\end{figure}

To ensure the probability of Type I error does not exceed $5\%$, a size analysis is also conducted. An equivalent experiment is conducted when $d = 0$, i.e. no separation exists between the two clusters, to determine size. See Figure \ref{power analysis} for results.

The estimated size was well below $5\%$. At each number of dimensions, the size experiment was simulated 100 times. Under the null hypothesis, the test never returned significant at 5, 10, and 20 dimensions. At 50 and 100 dimensions, the test returned significant only once each time. The test is conservative because the size must not exceed $5\%$ for any member of the composite null hypothesis.

\section{Application}

\subsection{Image Data Example}
\begin{figure}[H]
\begin{center}
\includegraphics[width=6in]{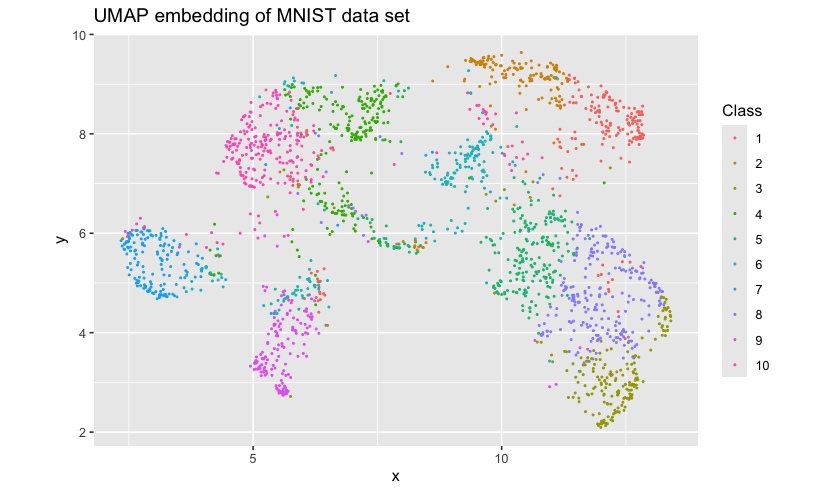}
\end{center}
\caption{UMAP embedding of the MNIST data set colored according to k-means clustering.}
\label{MNIST kmeans}
\end{figure}

To demonstrate use of the tool, we explore the MNIST data set in detail. The $(28 \times 28)$-pixel images were flattened and 1,500 samples were randomly sampled. A PCA pre-processing step was applied prior to applying UMAP \citep{UMAP} to construct a two-dimensional embedding. To replicate a real use case, i.e. when the true labels are unknown, we study a k-means clustering calculated on the high-dimensional data instead of the true digit labels (Figure \ref{MNIST kmeans}). The number of clusters, $k = 10$, was chosen to emulate the true digit labels. The reader may follow along using the \texttt{run\_example(example="MNIST", cluster="kmeans")} function in our \textit{DRtool} package.

At first glance, there are three major instances of disagreement between the UMAP embedding and the k-means clustering. Classes 1 and 2 seem to form one cluster together, class 4 is split into two separate clusters, and class 9 is merged with points from other clusters.

\subsubsection{Classes 1 and 2}
\begin{figure}[H]
\begin{center}
\includegraphics[width=6in]{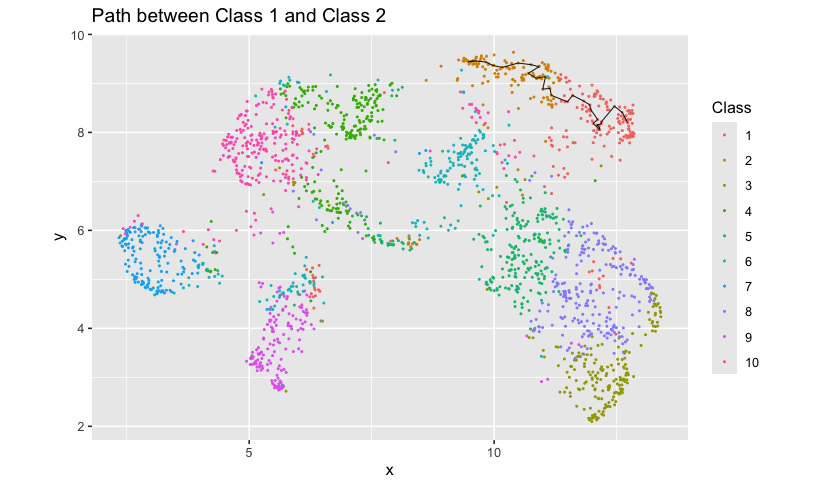}
\end{center}
\caption{Path between class 1 and class 2 in UMAP embedding of MNIST data set.}
\label{class 1+2 path MNIST}
\end{figure}

There seems to be minimal separation between classes 1 and 2, suggesting they may correspond to the same digit. We select a path from point 25,483 in class 1 to point 44,483 in class 2 (Figure \ref{class 1+2 path MNIST}). To get a closer look, we first look at the Path Projection Plot. The chosen number of dimensions is 100, which retains 97\% of the variance, and the path stabilizes at a CCA degree of four.

\begin{figure}[H]
\begin{center}
\includegraphics[width=6in]{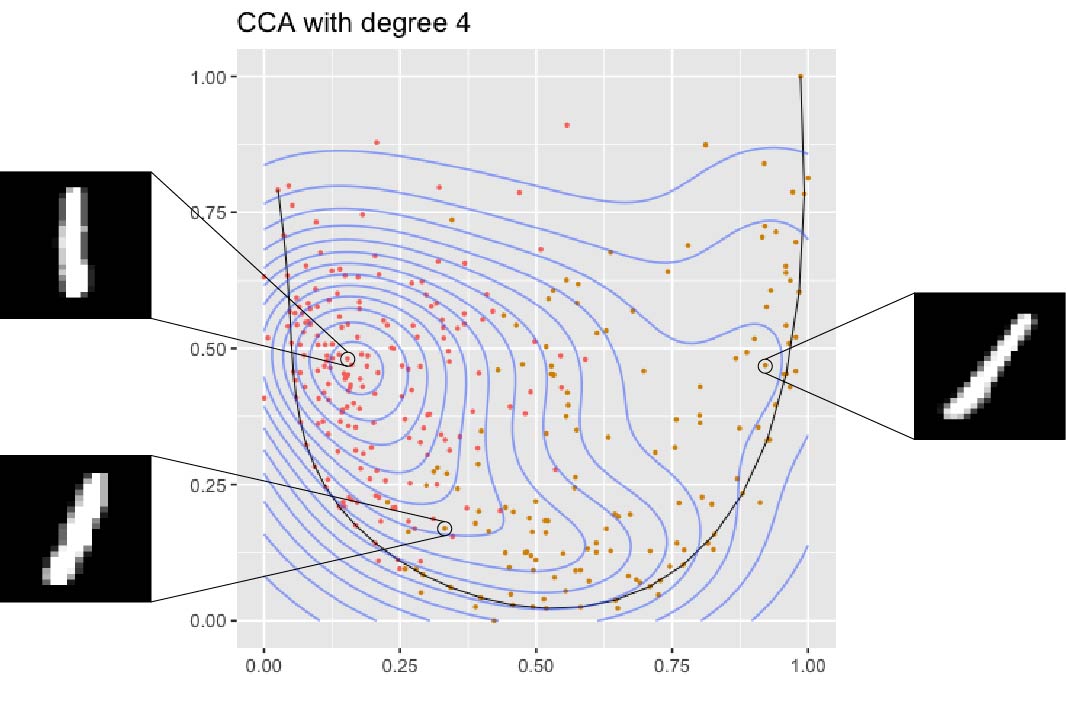}
\end{center}
\caption{Path projection plot of classes 1 and 2.}
\label{class 1+2 projection MNIST}
\end{figure}

The resulting plot depicts overlap between the two classes (Figure \ref{class 1+2 projection MNIST}). Adjusting the bandwidth of the density estimate to 1.5 shows unimodal density, suggesting the two classes may come from the same population. Showing the MST edges also does not provide any evidence of separation. The MST test results, however, may suggest otherwise. Seven crossings are counted when the approximate expectation under the null is 11.03 with a standard error of 3.523. While the bootstrapped $p$-value of $0.06$ is not significant at the 5\% level, the closeness indicates a more careful examination is necessary.

Inspection of the handwritten digits themselves reveals an interesting trend. While the majority of samples from both classes depict the digit one, the angle of the stroke differs drastically between the two classes. Following the path from class 1 to class 2, the strokes become more slanted. Both the MST path and MST test were able to detect this phenomenon, even though the two classes technically corresponded to the same digit.

Overall, the analytical plots provide a deeper look into the situation. The images of the digit one follow a skewed unimodal density centered around those drawn with a vertical stroke. The tail contains those drawn with more slanted strokes, specifically strokes drawn from the top right to the bottom left. While UMAP correctly captured this cluster, the gradual decline in density associated with increasingly slanted one digits is better depicted in the Path Projection Plot.

\subsubsection{Class 4}
\begin{figure}[H]
\begin{center}
\includegraphics[width=5in]{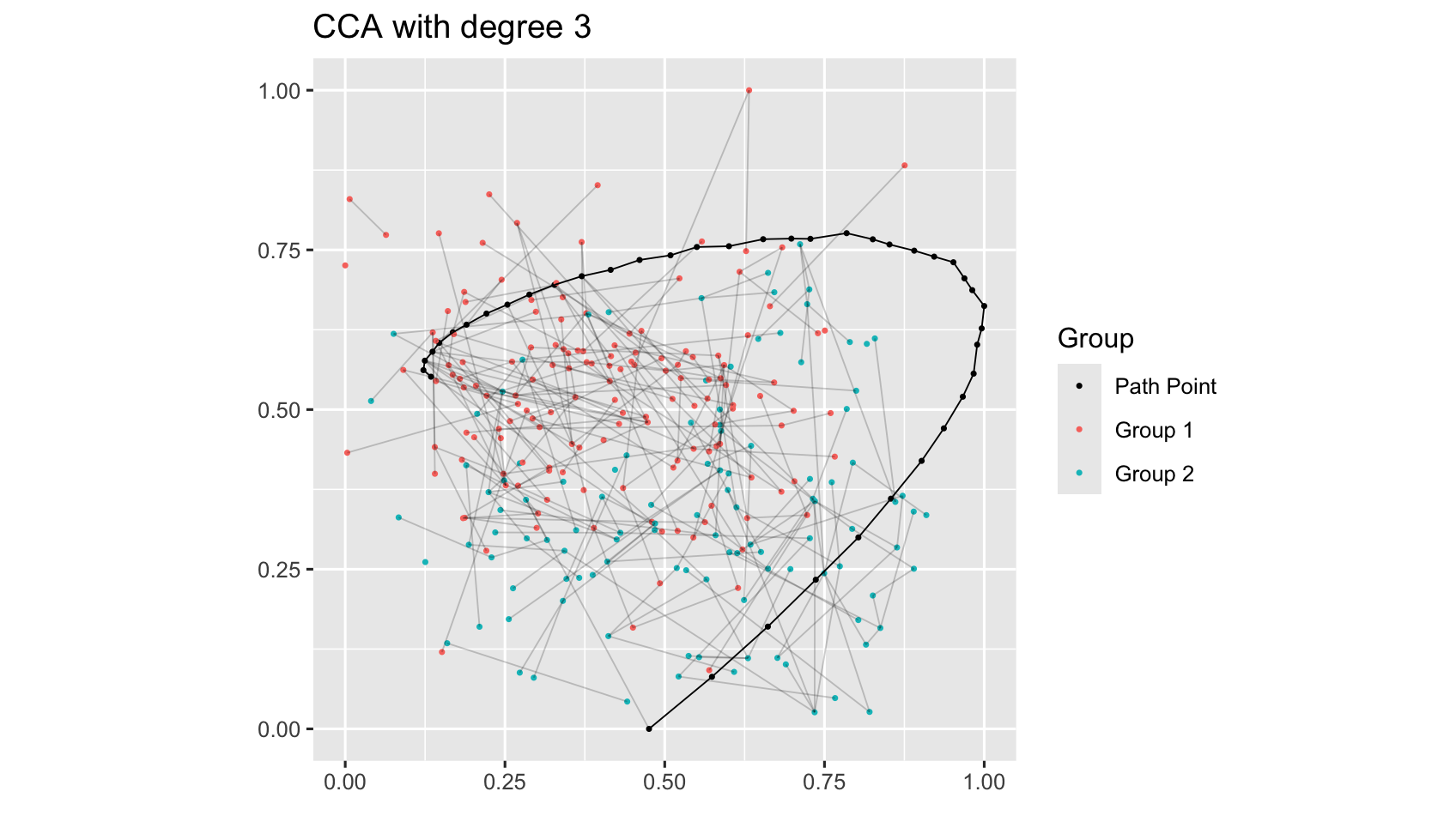}
\end{center}
\caption{Path projection plot of class 4 clusters.}
\label{class 4 projection MNIST}
\end{figure}

Class 4 is split between two different clusters in the UMAP embedding. We use the drawing tool to select the two clusters as our groups. The path projection settings are calibrated to 100 dimensions and a CCA degree of three. We also select the Group Coloring setting, so the points are colored according to group, rather than class. Analysis of the plot and estimated density does not provide evidence of separation. The MST edges, however, are more revealing after close inspection. If you look closely, there are few inter-group edges, even in overlapping regions (Figure \ref{class 4 projection MNIST}). By contrast, the MST test counts a larger number of edge crossings. 14 are counted when only 12.02 are expected with a standard error of 3.378.

According to the true class labels, these clusters correspond to distinct digits (Figure B1). The top class 4 cluster corresponds to the digit eight, while the bottom class 4 cluster corresponds to the digit five. The connecting class 1 cluster corresponds to the digit three. Three, five, and eight share common strokes, leading to blurred boundaries between their respective clusters and making it difficult for the MST test to detect. However, when increasing the sample size to 1,000 randomly sampled images of digits 3, 5, and 8 (up from 553), the MST test comes back significant for all pairwise comparisons (Table B1). Because the test is conservative by construction, the small effect sizes were difficult to detect at a smaller sample size. This is not particularly surprising because UMAP was also unable to detect the separation.

\subsubsection{Class 9}
\begin{figure}[H]
\begin{center}
\includegraphics[width=6in]{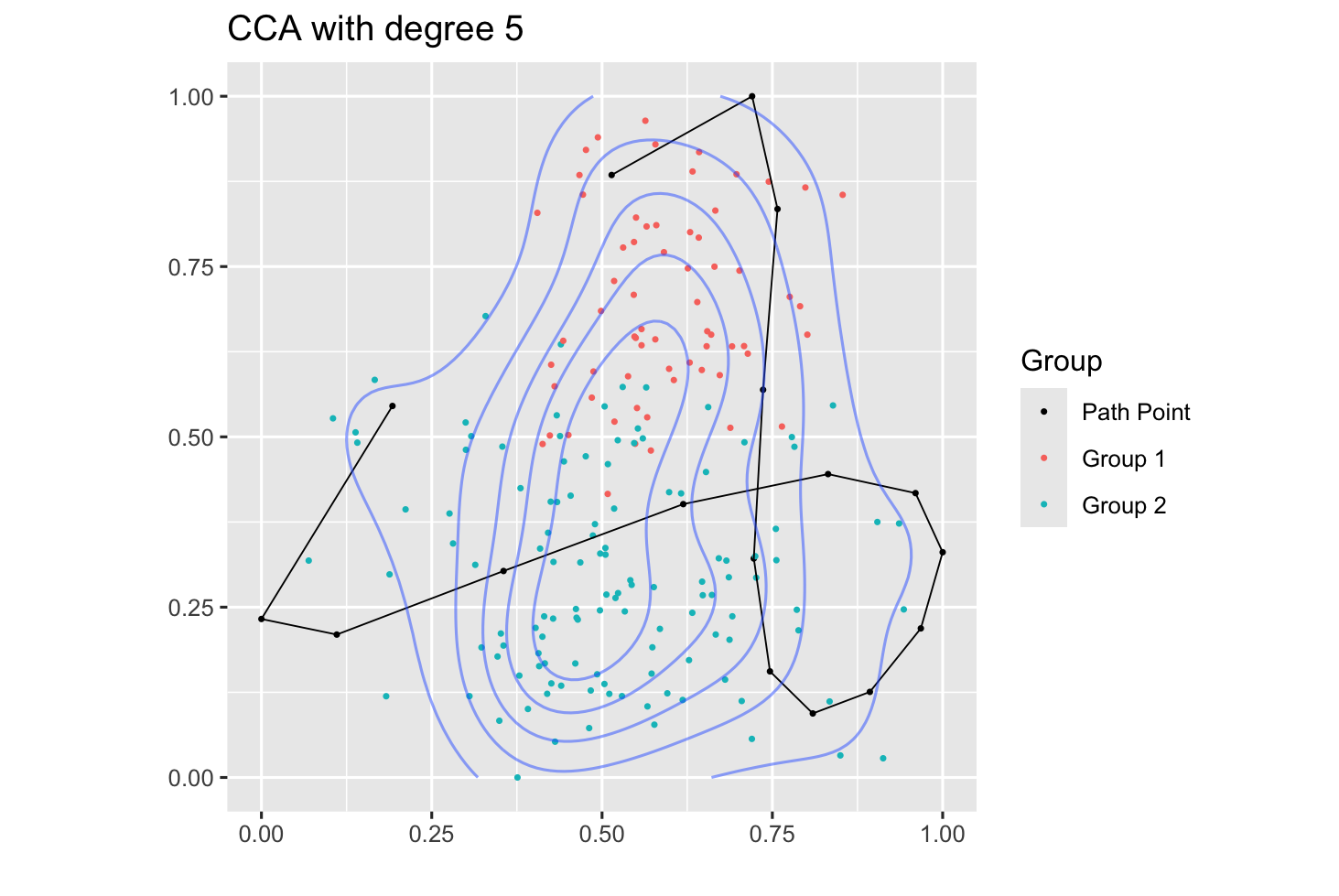}
\end{center}
\caption{Path projection plot of class 9 and remainder of cluster.}
\label{class 9 projection MNIST}
\end{figure}

Class 9 is well-separated, but its cluster also contains some points from other classes, mainly class 6. To determine if these points should belong to the same class, we use the drawing tool to select the class 9 points and the remaining points in the cluster as our groups. The path projection settings are calibrated to 100 dimensions and a CCA degree of five. Together, the points form a unimodal cluster, as shown by the approximate density calculated with a bandwidth of 1.3 (Figure \ref{class 9 projection MNIST}). Visually, there is also a consistent density of MST edges throughout the cluster, even where the two groups meet. The MST test agrees. There are 16 crossings counted, just below the expected value 16.37 under the null hypothesis. All evidence points towards the merging of these two groups.

According to the true class labels, this entire cluster corresponds with the digit six (Figure B1). The k-means clustering incorrectly scattered the points into multiple classes.

\subsection{Mass Cytometry Data Set}
\begin{figure}[H]
\begin{center}
\includegraphics[width=6in]{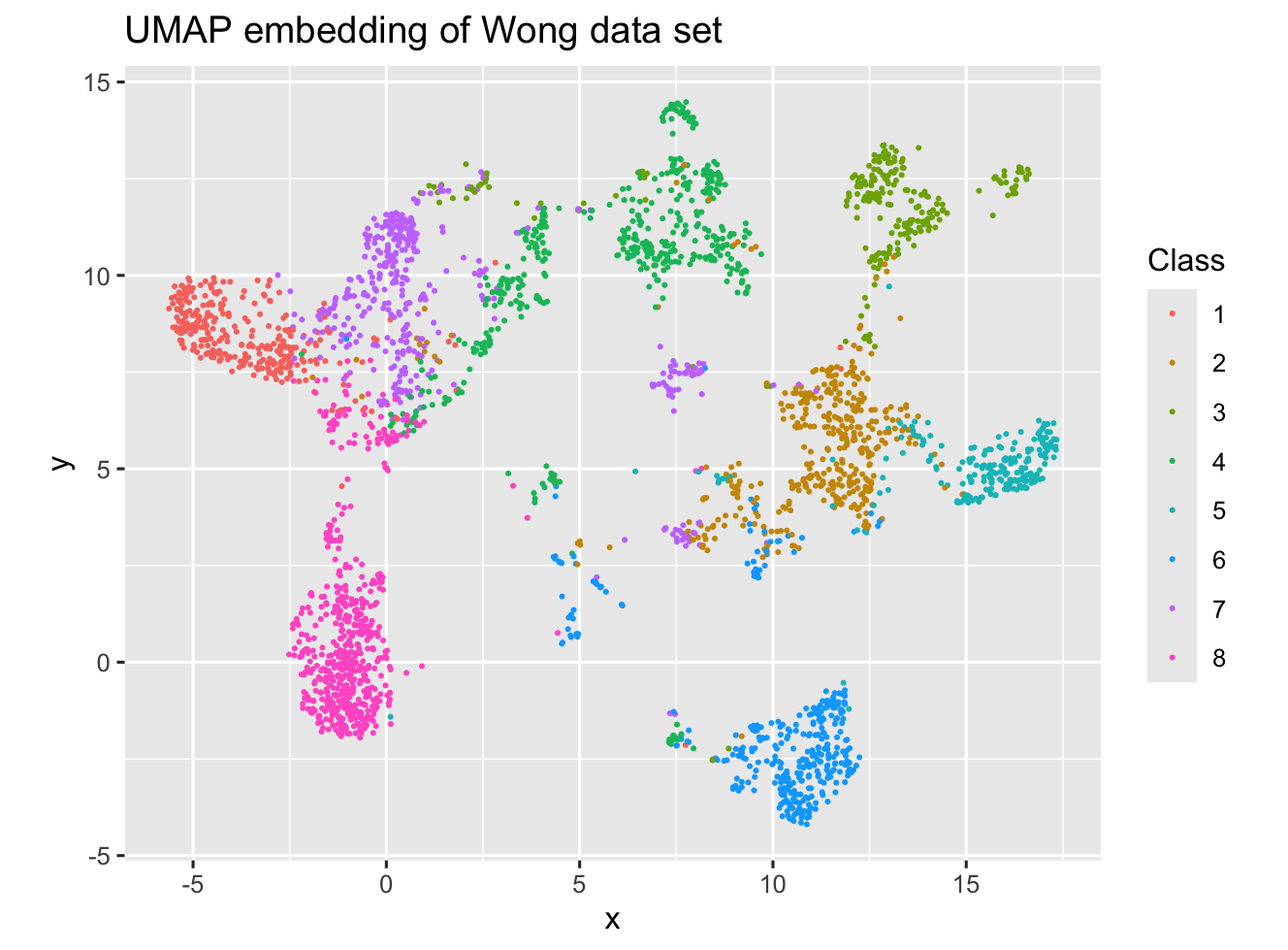}
\end{center}
\caption{UMAP embedding of the Wong data set colored according to k-means clustering.}
\label{Wong kmeans}
\end{figure}

We now explore a mass cytometry data set \citep{Wong_data_set} containing immune cell samples originating from various human tissues. The data was processed and the cells were classified into eight different human tissues according to the procedure used in \citet{UMAP_example}. Three thousand cells were randomly sampled. To replicate a real use case, we explore a $k$-means clustering calculated on the high-dimensional data, instead of the true class labels (Figure \ref{Wong kmeans}). The number of clusters, $k = 8$, was chosen to emulate the true class labels. The reader may follow along using the \texttt{run\_example(example="Wong", cluster="kmeans")} function.

Most of the $k$-means clustering seems to agree with the UMAP embedding. However, classes 4 and 8 are both split between two distinct clusters. Class 3 is also separated into three smaller sub-clusters.

\subsubsection{Class 4}
\begin{figure}[H]
\begin{center}
\includegraphics[width=6in]{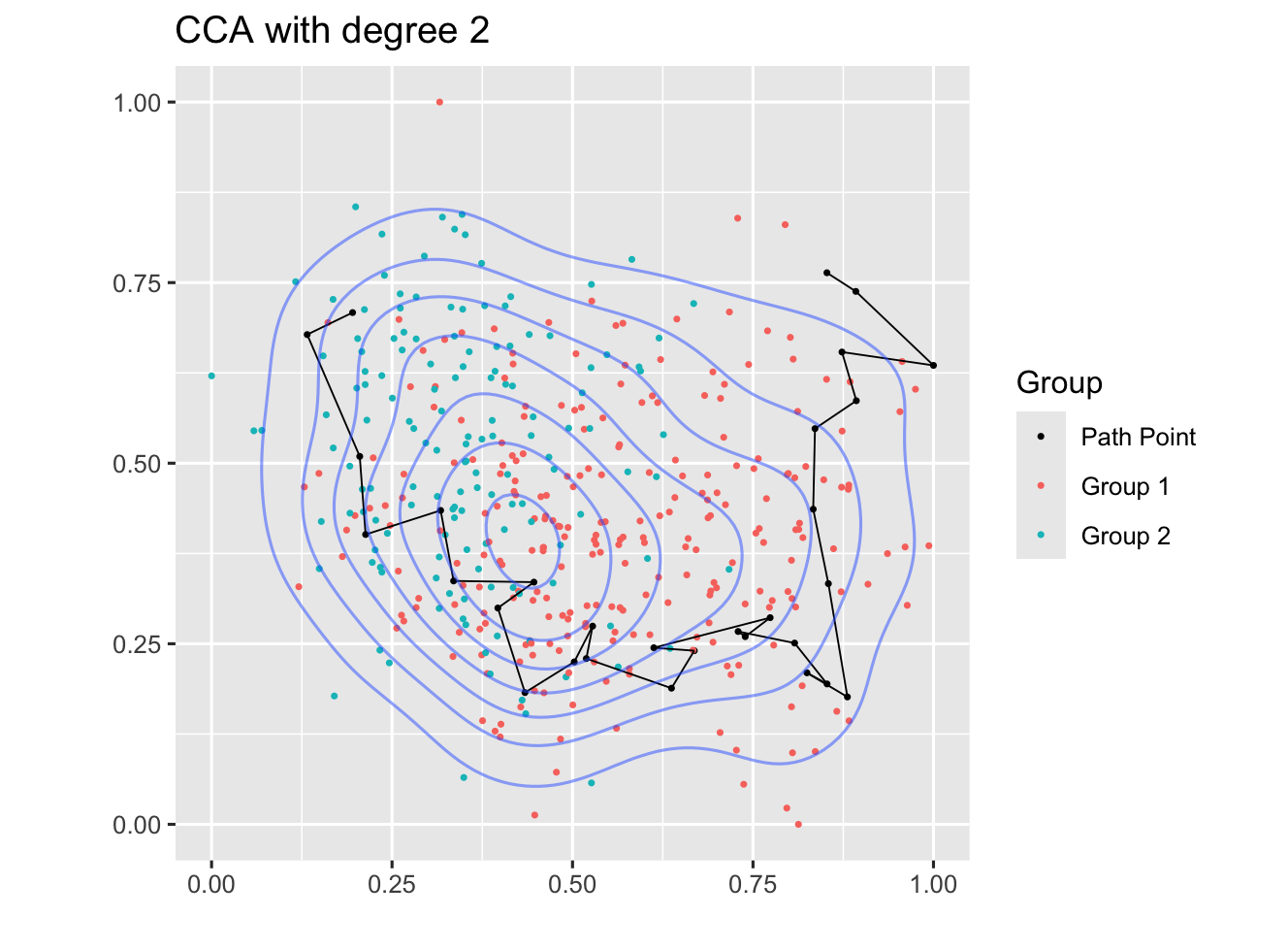}
\end{center}
\caption{Path projection plot of class 4 clusters.}
\label{class 4 projection Wong}
\end{figure}

Class 4 is split between two separate clusters in the UMAP embedding. To diagnose, we select the two custom clusters using the drawing tool. The path projection settings are calibrated to 20 dimensions and a CCA degree of two. We also select the Group Coloring setting, so the points are colored according to group, rather than class. The plot along with the estimated density does not provide any evidence of separation (Figure \ref{class 4 projection Wong}). The two selected groups also have 18 crossings, larger than the null expectation of 15.62. All evidence indicates the two groups were sampled from the same population, in agreement with the $k$-means clustering.

\begin{figure}[H]
\begin{center}
\includegraphics[width=6in]{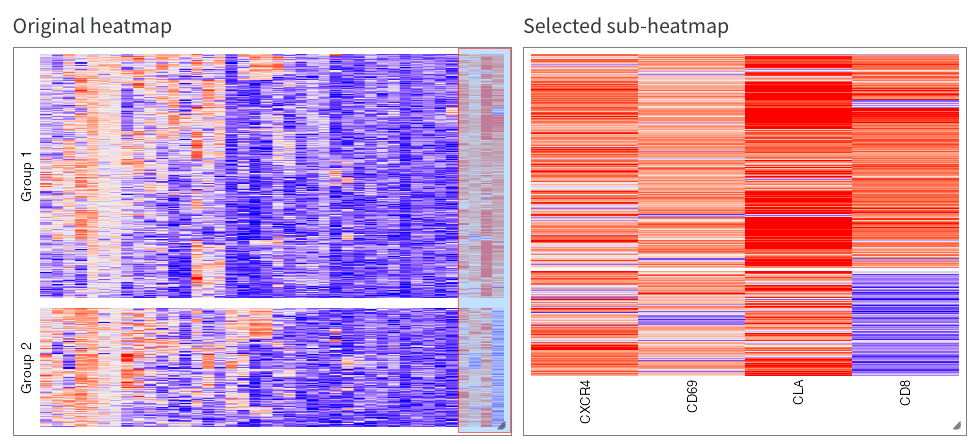}
\end{center}
\caption{Heatmap of cell counts within the two Class 4 sub-clusters.}
\label{class 4 heatmap Wong}
\end{figure}

To better understand why these two clusters are separated despite minimal evidence of separation, we reference the heatmap and nonlinear. According to the heatmap (Figure \ref{class 4 heatmap Wong}), the two groups differ most in CD8 protein counts. This is confirmed by the cell labels provided by \citet{UMAP_example}, which were passed to the tool as nonlinear (Figure \ref{class 4 meta data Wong}). So while separation was not observed by the MST, the discrepancy in CD8 protein counts accounts for the splitting of the class 4 points.

\begin{figure}[H]
\begin{center}
\includegraphics[width=6in]{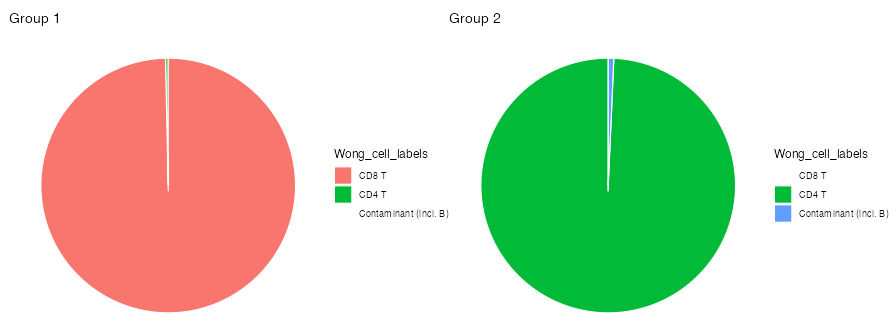}
\end{center}
\caption{Distribution of cell labels among cells in the two class 4 sub-clusters.}
\label{class 4 nonlinear Wong}
\end{figure}

It turns out that the $k$-means cluster got it right. Together, these two groups make up the cells sampled from skin tissue (Figure B2). However, there are two subgroups within the skin tissue cells differentiated by CD8 T cell count.

\subsubsection{Class 8}
\begin{figure}[H]
\begin{center}
\includegraphics[width=6in]{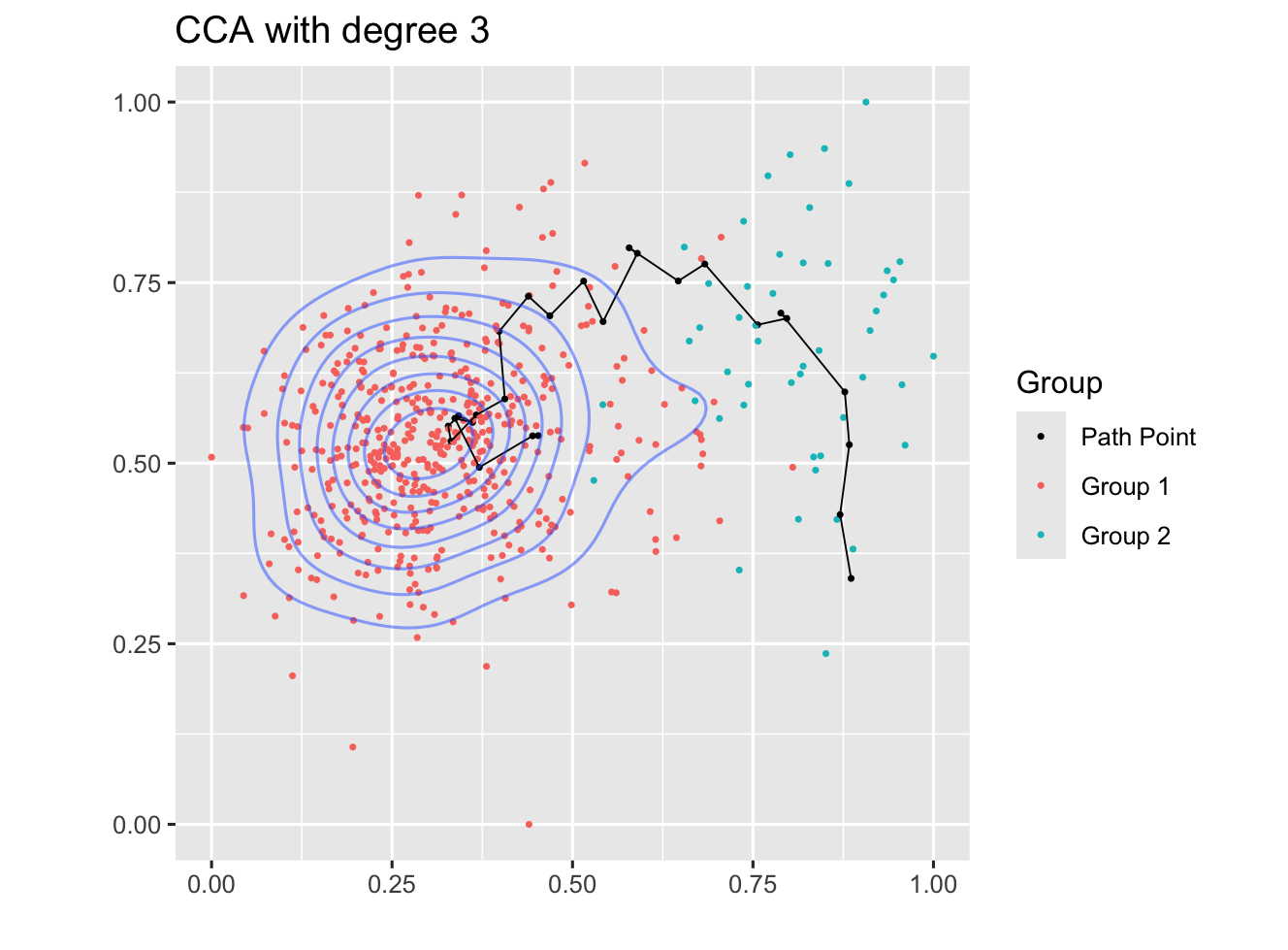}
\end{center}
\caption{Path projection plot of class 8 clusters.}
\label{class 8 projection Wong}
\end{figure}

The majority of class 8 points lie in a self-contained cluster. However, the rest lie in a separate nearby cluster. We select the two class 8 clusters as our two groups and study the projection of the path between them. The path projection settings are calibrated to 20 dimensions and a CCA degree of three. Similar to the UMAP embedding, the path projection plot depicts one dense cluster containing a majority of the points (Figure \ref{class 8 projection Wong}). The remainder of the points fall to one side in a low-density region. There is certainly not enough evidence to conclude the two groups belong to separate clusters from this plot alone.

Running the MST test returns interesting results. 20 crossings are counted when only 16.11 are expected with a standard error of 4.44. The number of crossings is greater than the number expected under the null hypothesis. This is due to the disparity in group densities. In order to preserve the size of the test, the null distribution must be constructed using a unimodal distribution whose density is similar to that of the less dense cluster (Section 3.4.2). This is why the expected number of crossings is so low. This phenomenon, however, has statistical meaning. The low relative density of group 1 provides minimal evidence from which we may draw conclusions. This lack of confidence is reflected in the low null expectation, and thus, the inflated p-value.

Overall, there is not enough evidence to declare the two groups are correctly separated. This is not quite correct according to the true labels. The two groups were not sampled from the same type of tissue, but the situation is slightly more complicated. See Figure B2 for details.

\subsubsection{Class 3}
The class 3 points are separated into three clusters -- two larger, elongated clusters to the left, and one smaller cluster to the right. Initial analysis of the path projection plot and MST test did not reveal any evidence of separation. However, the heatmap and nonlinear explain the separation captured by UMAP. The two larger clusters to the left are completely disjoint in their CD161 gene counts, but very similar in all other features. Meanwhile, the third smaller cluster is distinct from the other two larger clusters in its TCRgD counts. The nonlinear also reveal the left two clusters consist of CD8 T cells, while the right cluster consists of Tgd cells. 

To better understand why the MST was unable to capture the separation made apparent by UMAP, we investigated the original data. Recall that 3,000 cells were randomly sampled from a total of 327,457. Of these 3,000 cells, 242 of them belong to one of the two major sub-clusters in class 3. The MST test on these 242 cells did not reject ($p = 0.87$). To increase power, 4,000 cells were randomly sampled from class 3, and the MST test was run on those belonging to the two major sub-clusters. The test still did not reject, but was much closer at $p = 0.17$. When 5,000 cells were randomly sampled from cluster 3, the test did reject with $p < 0.01$. It seems the MST does capture the separation, but the test's power is too low to reject at smaller sample sizes.

\section{Discussion}
We have introduced our R package, \textit{DRtool}, and exemplified its use cases. The MST serves as an effective medium for understanding high-dimensional relationships and structures. The novel analytical tools provided by the package allow the user to extract the maximal amount of information from the MST by providing multiple perspectives. Such a multi-faceted view is necessary to understand contemporary dimension reduction methods that are trying to fit information from hundreds or even thousands of dimensions into only two dimensions. Advances in multiple fields have led to a surge in complex data, necessitating tools such as ours that help analysts assess and confirm their high-dimensional clustering results.

Further work should explore alternate methods for projecting paths into two dimensions. The goal of the projection is to ``unwind" the path, which is a nonlinear transformation, but nonlinear methods could pose two problems. First, most nonlinear methods do not have a natural out-of-sample extension that can be used to project points of interest other than the path points. Second, nonlinear methods can be prone to overfitting, especially when the path only contains a handful of points. On the other hand, linear methods define a linear transformation on the entire data space, so the projection naturally extends to points not on the path. Their rigidity also prevents overfitting. The downside is that linear methods are known to fail in high dimension due to the near-orthogonality of high-dimensional data. They also shrink space, which may obscure fine structural details that only nonlinear methods are capable of capturing.

Further work should also explore alternate methods of estimating cluster volume when calculating cluster density during the MST testing process. The product of singular values works well for clusters that are generally ellipsoidal or rectangular, but can fail for irregularly shaped clusters. A better estimate of the density could increase the power of the MST test.

\section{Code Availability}
All data and code are freely available at \url{https://www.github.com/JustinMLin/DRtool}.

\section{Declaration of Interest}
The authors report there are no competing interests to declare.

\section{Supplementary Material}
The proof for the null hypothesis problem presented in Section 3.4.2, Figures B1/B2, and Table B1 are provided.


\clearpage
\bibliographystyle{chicago}
\bibliography{ref}

\end{document}